# The Sub-band Vectoring Technique for Multi-Operator Environments

Francesco Vatalaro, *Senior Member, IEEE*, Franco Mazzenga, Romeo Giuliano

*Abstract*—Self- and alien-FEXT disturbances on the same cable severely limit performance in the commonly employed VDSL2 17a, i.e. 17.6 MHz, standard profile. Then, in a multi-operator environment, VDSL2 may be unable to provide the 100 Mbit/s speed required by the EC's policy target, unless vectoring is adopted along with a suitable Multi-Operator Vectoring (MOV) technique. Some vendors recently proposed enlarging the bandwidth up to 35.2 MHz (so-called e-VDSL) as one possible solution to increase the 100 Mbit/s coverage. However, as we show in the paper, the bandwidth advantage is illusory as an "ideal MOV" is not implementable with today's technology, and alien-FEXT practically destroys the envisaged data-rate increase. Therefore, we introduce the Sub-Band Vectoring (SBV) technique, as one practical solution, which turns out to be near optimal in terms of achievable data-rate, as well as incremental for those Countries where the NRA adopts sub-loop unbundling regulation. SBV adopts a novel frequency division multiplexing approach, which imposes one fairness condition to equalize data-rate values and overall bandwidth per operator at different distances. The paper shows that e-VDSL with SBV is robust, as it can cope with relatively high values of vectoring impairments. It also shows that the SBV allows up to three co-located operators with e-VDSL to coexist, without imposing any of the burdensome coordination limitations of the ideal MOV architecture. Our results show that the SBV allows achieving up to 210 Mbit/s per user for e-VDSL and up to 620 Mbit/s per user for a bandwidth expanded up to 105.6 MHz, with two telecom operators. Finally, the paper shows that a migration strategy towards usage of the G.fast standard thanks to SBV is achieved in a simple way.

*Keywords*—Vectoring, VDSL2, e-VDSL, SBV, Sub Loop Unbundling, FEXT, Multi-Operator Vectoring

I. INTRODUCTION

Worldwide, in most OECD Countries – and the EU is no exception – serious efforts are ongoing to modernize the telecommunications access networks. There are generally two concurring thrusts in such direction. The first thrust is increase of demand for new services, especially for residential and SOHO (small office - home office) applications, which are gradually appearing, such as 4K TV and unicasting Internet services, Cloud computing and the Internet of Things. The second but not less important thrust are the UBB (Ultra Broad Band) policy programs underway in most Countries targeting year 2020, such as "Connecting America: The National Broadband Plan" in the U.S. and the European Commission's "Digital Agenda for Europe" (DAE) programme, which was set up in 2010.

Solutions such as FttH (Fiber-to-the-Home), and similar architectures, appear the best choice for those Countries with poor quality pre-existing copper loops in the access network, and – in more developed Countries, too – for those green-field conditions where municipalities established urban plans for new built-up areas or deeply renovated areas. Except for those conditions, FttH may lend itself to uneconomical and risky implementations, which often discourage telecom operators to dismiss quickly their legacy copper network towards a full-optical access.

Therefore, operators mostly seek for hybrid fiber-copper solutions capable to improve performance on their existing networks, so that investments towards the ultimate FttH solution can be gradual and safe, while at the same time achieving, and tracking, the needed level of customer satisfaction. Gradual performance improvement in their copper access networks has been always the preferred telecom operators' strategy, generally based on continuous improvement of the DSL (Digital Subscriber Line) family of standards e.g., ADSL, ADSL2+, and VDSL2.

The FttC (Fiber-to-the-Cabinet) architecture emerged as an important enabler to provide UBB in most European Countries. Currently, thanks to FttC telecom operators are able to provide their customers in large coverage areas with 30 Mbit/s, or so, data-rates by using the VDSL2 (Very high-speed Digital Subscriber Line type 2) profile 17a technology (ITU-T G.993.2), having 17.6 MHz bandwidth.

However, operators aim at significantly improving the data-rate in the near future, possibly well beyond 100 Mbit/s, while at the same time allowing gradual networks deployment closely following service demand. Therefore, two of today's most promising solutions are:

- The "Enlarged-bandwidth VDSL" FttC access network infrastructure, sometimes referred to as e-VDSL or "Vplus" [1], especially useful for heavily urbanized Countries, e.g. Italy, with short copper loops (250 m, or so) between the street cabinet (CAB) and the condominiums' Distribution point (Dp).

- The G.fast (Fast Access to Subscriber Terminals) ITUstandard (ITU-T G.9701) aiming at bringing the fiber up to the Dp – the so-called FttDp architecture – especially useful for medium-to-light density urbanizations having long CAB-to-Dp distances and clustered condominiums attached to one, or a few, Dps.

F. Vatalaro, F. Mazzenga are with the Department of Enterprise Engineering "Mario Lucertini", University of Rome Tor Vergata, Via del Politecnico 1, 00133, Rome, Italy, *vatalaro@uniroma2.it*, *mazzenga@ing.uniroma2.it*
Romeo Giuliano is with the Department of Innovation & Information Engineering, Guglielmo Marconi University, Via Plinio 44, 00193, Rome, Italy, *r.giuliano@unimarconi.it* (contact author).

Investment costs for FttDp solutions are generally much higher than costs needed for FttC architectures. Therefore, at present FttC architectures are preferred when implementation conditions are favorable, while at the same time they can also meet the UBB policy targets.

Self- and alien-FEXT (Far End Cross Talk) disturbances which emerge on the same cable among user signals of the same or different telecom operators respectively, severely limit performance in the commonly employed VDSL2 17a, i.e. 17.6 MHz, standard profile as well as for e-VDSL and G.fast. As we verified by extensive computer calculations, the current VDSL2 17a standard profile is unable to meet the 100 Mbit/s DAE data-rate target on the Italian copper access network.[1] Even in the fifteen largest Italian cities where CABs' distribution is dense and copper line lengths are among the shortest ones, achievable speed under medium-to-heavy traffic loads is roughly limited to about 30 to 50 Mbit/s, and the datarate may be highly unstable due to FEXT.

Regulatory remedies such as the SLU (Sub Loop Unbundling), which aim at promoting infrastructure competition [3] can make usage of the so-called "vectoring" technology (ITU-T G.993.5) essentially useless at the CAB. To solve this problem, some National Regulatory Authorities (NRAs) are considering Multi-Operator Vectoring (MOV) systems [4]. However, these solutions are hardly feasible in a short time and without imposing large cost penalties to operators, having also to strictly coordinate their deployments at every addressed CAB. The "ideal MOV" presents several burdensome limitations. Some of them are as follows. MOV is not compatible with already installed technologies and systems. It faces obstacles related to the lack of an international standard for ensuring interoperability between different vendors equipment. In addition, the equipment is complex, not always easy to install, and can be very power-hungry thus leading to the increase of operating costs and to significant impact on the ecological footprint. Last but not the least, it may present privacy issues between subscribers of different operators. Table I provides a qualitative comparison between the "ideal MOV" configuration and the technique proposed in this paper.

In devising a practical MOV solution, the following constraints should be jointly considered:

- the present spectrum usage pattern up to 17.6 MHz should not be modified, since it is standardized and telecom operators are already using it;
- higher frequencies can be considered to increase the data-rate, however the migration path towards G.fast should be taken into account to allow future data-rate growth;
- the presence of multiple telecom operators having their proprietary infrastructure without need of strict coordination between them should be promoted;
- minimum, or possibly none, technological modifications should be imposed, to speed up implementations without going through long standardization and deployment phases.

Given the ideal MOV difficulties and limitations, some main vendors have proposed solutions for enlarging the bandwidth up to 35.2 MHz of the copper pair to obtain increased datarate even without vectoring at the CAB. Such solutions are promising also by virtue of the possible use of powerful channel codes, which lend themselves to reaching extension for one single operator well beyond the typical lengths of VDSL2 with vectoring – e.g., 400 Mbit/s within 300 m and 100 Mbit/s within 800 m [5]. However, they do not address the problem of alien-FEXT due to two or more co-located operators in the presence of distinct vectoring groups.

Therefore, due to limitations of present and forthcoming equipment in a multi-operator scenario in this paper we propose a new technique, the Sub-band Vectoring (SBV) technique, as a "practical MOV" to be engineered and implemented in a short time. The SBV concept is based on frequency division multiplexing and allows vectoring use without sacrificing SLU, while it does not require complex coordination nor synchronization between co-sited telecom operators.

In particular, in this work we propose to use high frequen-

TABLE I. COMPARISON BETWEEN THE CHARACTERISTICS OF THE IDEAL MOV AND THE SBV TECHNIQUE PROPOSED IN THIS PAPER.

| Feature | Ideal MOV | SBV |
|---|---|---|
| International standard | Required | Advisable but not required |
| Spectrum efficiency | Maximum | High |
| Number of operators | 4 (AGCOM* spec) | 3 for 35.2 MHz bandwidth, 4 or more for increased bandwidth |
| Equipment installation | Complex | Easy |
| Privacy concerns | Yes | No |
| Power consumption | High | Very low |
| Ecological footprint | Can be high | Low |

* Autorità per le Garanzie nelle Comunicazioni (AGCOM) is the Italian NRA.

cies to increase the offered data-rate. The selected frequency band(s) are above the standardized 0-17.6 MHz in order to ensure backwards compatibility with VDSL2 17a profile. Higher bands are first partitioned into separated band slots to be exclusively assigned to each telecom operator. In each assigned slot, the single operator is free to adopt own selected technology without incurring in alien-FEXT. Moreover, vectoring techniques can be used to cancel out self-FEXT in an assigned slot.

This paper is organized as follows. In Section II we introduce the proposed SBV solution. In Section III the SBV technique is analyzed in detail considering different aspects. After having introduced the performance figures and propagation and interference models, first we highlight the rationale for separating the bands between telecom operators; then we analyze the problem of the identification of the band slots to be assigned to operators. The resulting band partitioning scheme is selected to ensure fairness among operators. Other SBV performance results concerning the extension of the transmission band above 35.2 MHz, have been presented. The Section concludes with the evaluation of the SBV performance degradation due to non-ideal vectoring operations. In Section IV the compatibility with G.fast is discussed. Finally, in Section V we report on our conclusions.

---

[1] We did calculations massively (i.e. 100%) for the cities classified by the Italian Government as Cluster A area, and in a very large percentage (slightly less than 80%) in the Cluster B area. Cluster A collects the 15 largest Italian cities (almost 10 million inhabitants), while Cluster B collects about 1200 large-to-medium cities (about 27 million inhabitants) [2]. Overall Cluster A and Cluster B population is more than 60% of the Italian population.

## II. THE SBV PROPOSAL

We assume DSL signals of different telecom operators originate from the same, or adjacent, street cabinet(s).[2] In order to keep backwards compatibility with already existing VDSL2 systems in the 0–17.6 MHz band, we retain telecom operators continue sharing this band in accordance with regulatory rules specific for each Country. To fix ideas, as is in Italy vectoring is not applied in the 0–17.6 MHz band, in which the number of VDSL2 operators is unrestricted. We assume the presence in the same CAB area of different DSL standards and, in our modeling, we will show how coexistence between VDSL2 and e-VDSL can be ensured.

Let the frequency band above 17.6 MHz, used by e-VDSL and, in future, by possible new bandwidth-enhanced DSL standards, is partitioned into multiple "sub-channels" each having 17.6 MHz bandwidth, as depicted in Figure 1, $i = 0,1,…M − 1$. In the case of e-VDSL, M = 2, while in this paper we account for values up to M = 6, i.e. frequencies up to 105.6 MHz. To keep our analysis simple and effective, we extend the bandwidth planning above 17.6 MHz only assuming increase in downstream channels demand. In other words, no additional band is allocated to upstream traffic. Again, this is only to fix ideas, as the proposed SBV technique does not limit upstream allocations, in any respect. In fact, the actual frequency planning for the upper bands will be a task for the NRA based on characteristics and balance of the expected traffic requirements. This is out of the scope of this paper. In the frequency spectrum above the 0–17.6 MHz "sub-channel 0", the bandwidth of each sub-channel is further divided into consecutive assignment band slots having bandwidth $B_k$, $k = 1,2,…,K$. These slots are partitioned in accordance with some (not necessarily regular) pattern and assigned to the $N_{op}$ telecom operators, possibly according to rules issued by the NRA.

The case depicted in Figure 1 refers to $N_{op}$ = 2. For a given value of $N_{op}$, the selection of the bandwidth $B_k$ is of importance to ensure fair resource assignment among operators. The proposed band slot partitioning is discussed in Section III-C.

Figure 2 shows one possible reference architecture for the SBV transmit unit based on the band partitioning shown in Figure 1. The SBV signal to be transmitted on the generic sub-channel is generated at baseband using existing VDSL2 modulators, which are designed (and optimized) for the 0–17.6 MHz band. Figure 2 refers to the case where the 0–17.6 MHz channel is possibly non-vectored. The baseband signals corresponding to the upper sub-channels are frequency translated from baseband to the proper center frequency by using a single side band (SSB) modulator and sub-sequent up-conversion unit. Each operator independently transmits on the higher channels by only modulating their assigned set of DMT (Discrete Multi-Tone) carriers in each of the slots. To counteract self-FEXT effects, one telecom operator can decide to implement vectoring, as well as any kind of signal processing, on the tones assigned to it.

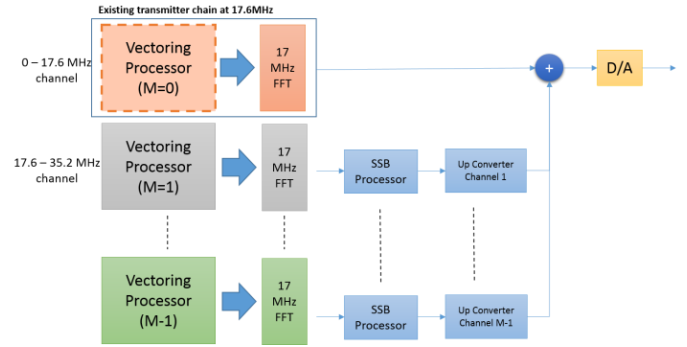

Fig. 2. SBV reference transmitter architecture.

If backwards compatibility is not required, partitioning could be extended to the 0–17.6 MHz channel. This is generally a choice of the NRA. One example is when the NRA did not yet impose sector-specific regulation, and telecom operators did not yet implement a VDSL2-FttC architecture in the Country or in a geographical area.

Several practical implementations can be proposed for the SBV DSL-access multiplexer (SBV-DSLAM) [6]. While in some architectures each operator installs own proprietary equipment, in other ones only one operator (generally the incumbent operator) installs and runs one SBV-DSLAM allowing one kind of frequency unbundling to competitors.

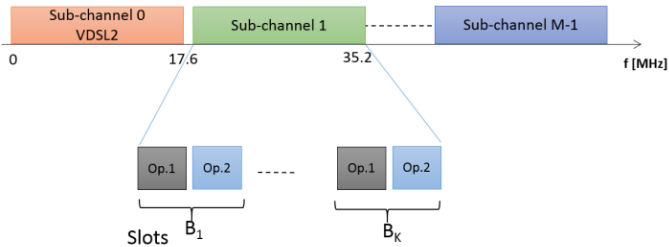

Fig. 1. Considered bandwidth partitioning for the SBV solution.

---

[2] In this paper, we do not consider compatibility issues with ADSL and ADSL2+, while we concentrate on those issues for the VDSL2 family of profiles and G.fast with the forthcoming e-VDSL standard, as well as possible alternatives with further extended bandwidths.

III. SBV ANALYSIS

In this Section we investigate the effectiveness and the performance of the SBV technique considering some different and related aspects and variable interference scenario conditions. The first and important aspect is to assess under what conditions the SBV technique overcomes simple non vectored (NV) shared use of the full bandwidth. The second is related to the problem of sub-band partitioning among operators. This aspect has a strong impact on the actual implementation of the SBV so that users served by different telecom operators experience similar performance under the same channel conditions. The application of SBV technique for bands up to 105.6 MHz and the comparison of achievable performance with the NV case is also investigated. Finally, we discuss the SBV performance in the presence of vectoring imperfections. In all cases, the analysis is carried out using computer calculation, considering the same scenario and adopting commonly used models for propagation and cross-talk interference.

*A. Performance parameters, channel and interference modelling*

For comparison purposes, the expressions for the aggregate data-rate of the full shared band NV case and of the dedicated partial band (i.e., SBV) vectored case, for the generic *i*-th user are, respectively:

$$R_{b,NV}^{(i)} = R_s \sum_{k \in S_{NV}} \log_2 \left( 1 + \frac{|H_d^{(i)}(k)|^2 \cdot P_{k,NV}^{(i)}}{\left(\eta_k^{(i)} + I_k^{(i)}\right) \Gamma} \right) \quad (1)$$

$$R_{b,SBV}^{(i)} = R_s \sum_{k \in S_{SBV}} \log_2 \left( 1 + \frac{|H_d^{(i)}(k)|^2 \cdot P_{k,SBV}^{(i)}}{\eta_k^{(i)} \cdot r_{V,k} \cdot \Gamma} \right) \quad (2)$$

where (omitting the dependence on *i* in the notation) we have:

- $R_s$ is the symbol rate, i.e. $R_s$ = 4 ksymb/s for tone frequency spacing $\Delta f$ = 4.3125 kHz;
- $S_{NV}$ and $S_{SBV}$ are the sets containing the indexes of the tones selected for the reference user in the NV case and in the SBV case, respectively;
- $\Gamma$ = 12 dB [7];
- $P_{k,NV}$ and $P_{k,SBV}$ are the powers allocated to the *k*-th tone for NV and SBV, respectively;
- $I_k$ is the FEXT term on the *k*-th tone;

- $H_d(k)$ is the direct propagation term depending on the cable type and on the distance between transmitter and receiver [8];
- $r_{V,k}$ is the vectoring degradation factor, which takes into account the non-ideal interference cancellation of the vectoring technique;
- $\eta_k$ is the background noise power accounting for the noise power spectral density $N_0$ = -140 dBm/Hz.

Even though not evidenced in (1) and (2), bit loading limitations between 2 and 15 bits per symbol have been applied in the calculation of the data-rates. For simplicity, in the following we assume flat spectral transmission power both in 0 – 17.6 MHz and in the higher bands, i.e. no bit loading algorithm was considered. Moreover, each telecom operator transmits the maximum available power according to the international standard. Thus, we have $P_{k,SBV} = N_{op} \cdot P_{k,NV}$. The generic FEXT term in (1) is:

$$I_k^{(i)} = \sum_{m=1, m \neq i}^{N_{us}} \left| H_{F,k}^{(i,m)} \right|^2 \cdot P_{k,NV}^{(m)} \quad (3)$$

where $N_{us}$ is the number of interfering signals. The term,

$$H_{F,k}^{(i,m)} = H_{FEXT}(f_k, d_i, l_{i,m}) \quad (4)$$

is the FEXT channel transfer function evaluated at tone frequency, $f_k$. It depends on the distance between the street cabinet and the *i*-th user, $d_i$, and by the coupling length between the *i*-th user with the *m*-th interferer, $l_{im}$. We assume a statistical fluctuation of $H_{FEXT}(f,d,l)$ with respect to the FEXT contribution non-overcome for the 99% of the cases, $H_{FEXT,99}(f,d,l)$:

$$H_{FEXT}(f,d,l) = H_{FEXT,99}(f,d,l) \cdot e^{-j\phi(f)} 10^{-\frac{X_{dB}}{20}} \quad (5)$$

where $\varphi(f)$ is an irrelevant phase term and $X_{dB}$ is the FEXT fluctuation which, in accordance with [9], is assumed to be Gaussian with mean 11.65 dB and standard deviation 5 dB. According to [9] this is the high coupling FEXT condition, due to the interfering pairs close to the reference pair in the same binder. In our calculations, we also assume the LQ-Gamma cable whose propagation characteristics are referred in [10].[3]

The $H_{FEXT,99}(f,d,l)$ model used for our calculations is that provided in [10], which was experimentally tested up to 300 MHz. However, to further assess the validity of this model we made comparisons with other classical FEXT models [9], [11] and we observed that, when only intra-binder interference is considered (see below), model in [10] is more conservative than extending classical models beyond their tested range.

To further reduce alien-FEXT, telecom operators could enable vectoring in the 0 – 17.6 MHz band to cancel self-FEXT.

---

[3] Direct propagation and FEXT models for frequencies higher than 30 MHz are available for G.fast only and for street cabinet to network termination distances up to 200 m. For this reason the selection of these two parameters could be a critical step. From simulations we observed that data-rate performance obtained for the LQ-Gamma cable model in [10] are close to those obtained using the AWG24 cable (typically used in Italy) at least up to 17.6 MHz. To the best of authors' knowledge, the AWG24 channel model for frequencies up to 105.6 MHz is not currently available in the literature.

Obviously, reduction/cancellation of self-FEXT in the 0–17.6 band is beneficial for the NV solution and it also improves performance of the SBV solution presented in this paper.

## B. Why band partitioning can be effective?

In this subsection and in the following one we address two related problems. The first problem is assessing under what conditions the SBV technique overcomes simple NV shared use of the full bandwidth. The second problem is how to actually implement SBV so that users served by different telecom operators experience similar performance under the same channel conditions.

To justify the effectiveness of band partitioning and the SBV strategy we propose in practical settings, we must compare it with non-vectored full bandwidth usage. In fact, ideal MOV promises optimal transmission performance because for all operators using vectoring the channel conditions are AWGN (Additive White Gaussian Noise) on the full available band. However, considering the above-mentioned MOV feasibility limitations, it is interesting to evaluate performance with band partitioning, aiming at identifying the maximum number of operators that can provide a data-rate objective (e.g., 100 Mbit/s), in spite of the reduced allocated resource. We now aim at comparing the performance achievable with vectoring on the set of bands assigned to one operator and the performance achievable on the full shared bandwidth without vectoring. In fact, in the copper pair both the attenuation and FEXT strongly increase with frequency. Therefore, in such a channel the signal-to-overall disturbance, SNR, rapidly decreases beyond a frequency, typically in the range 10 – 12 MHz, which actually depends on operational conditions. Then, the bandwidth extension beyond 17.6 MHz can be little effective, or even ineffective, if the AWGN condition is not met. In our calculations restricted to the 35.2 MHz band, we have considered the band slots shown in Table II. For frequencies below 17.6 MHz, band slots are equal to the bands indicated in profile 17a for downstream transmissions. For higher frequencies each band slot contains about 1024 DMT tones corresponding to about 4.5 MHz. Band slots have been (ideally) partitioned by assigning tones to operators in an alternate manner, i.e. one tone to one operator and the next tone to another one and so on until all the tones in the band slot have been assigned.[4] Finally we assume a co-located scenario, in which the reference and the $N_{us}$ interferers are located in the same building and belong to the same binder. The additional distance due to subscribers being located in different building floors has been neglected in our calculations. Furthermore, depending on the cable structure, one binder can typically contain 25 or 50 copper pairs. In our calculations only intra-binder interference has been considered, i.e. an optimistic scenario for NV has been considered. In Figure 3 we show the achievable data-rates for the single operator as a function of the bandwidth number defined in Table II, for the NV case (solid lines) and the SBV case (dashed lines) with $N_{op}$ = 2, 3 and a distance between the street cabinet and the Network Termination (NT), i.e., CAB-to-NT distance, d=100 m (case a and case c) and d=250 m (case b and case d), respectively. Data-rates refer to the 10-th percentile and have been obtained assuming a vectoring degradation factor $r_V$ = 10 dB, constant for all tones. The number of active interfering users, $N_{us}$, in the binder has been varied from 2 (very low load), 6 (low load), 12 (medium load), to 24 (high load). These choices only impact performance in the NV case. The aggregated data-rate per operator can be obtained by summing the achievable datarates in each bandwidth numbered from 1 to 7, both for NV and SBV.

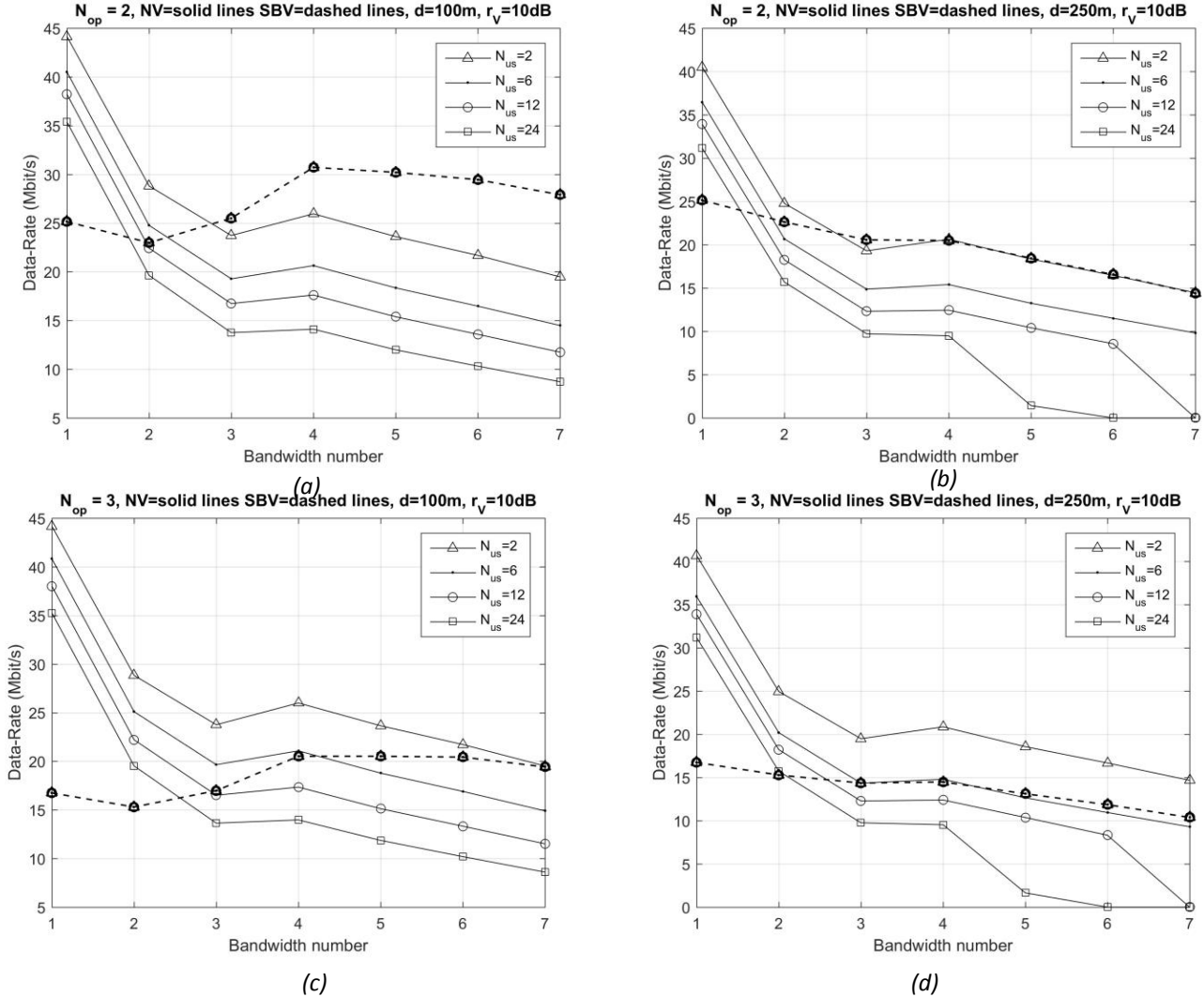

Fig. 3. Comparison of the data-rates in each bandwidth for the NV and the SBV cases: $N_{op}$=2, CAB-to-NT distances of 100 m (case a) and 250 m (case b), respectively, and $N_{op}$=3, CAB-to-NT distances of 100 m (case c) and 250 m (case d), respectively. Bandwidth numbers refer to frequency ranges in Table II.

Results show that the data-rate in the SBV case is almost constant, and decreases with the CAB-to-NT distance and with frequency, due to direct channel propagation. As expected, in the NV case performance degrades with the number of active interfering users and for high frequencies due to the greater FEXT contribution. For frequencies lower than about 10 – 12 MHz, FEXT contribution is negligible, so NV overcomes SBV in the 0 – 17.6 MHz bandwidth. Consequently, for a lower frequency band it is not convenient to allocate separate bands to the assumed three telecom operators. On the contrary, for higher frequencies (above 12 MHz, or so), SBV overcomes NV. Therefore the SBV technique improves DSL performance above a certain frequency threshold in a multioperator scenario, without any further needs of coordination. It is noteworthy that, when $N_{op}$=2, SBV overcomes NV already above 5 MHz for all distances of interest under low to medium-high interference conditions. As expected, under a very low load condition ($N_{us}$ = 2) NV overcomes SBV performance for $N_{op}$ = 3. However, very-low-load conditions do not represent a mature deployment scenario, so that they should not be considered as indicative when planning the DSL access system. In particular, a NRA should not consider lowload conditions as a reference, in order to avoid complex bandwidth reallocation procedures during the system lifespan. As a preliminary conclusion, we see that, when the NRA already took the decision to allow operators to compete on a lower band (typically, 0 – 17.6 MHz), there is no need to change their regulation in such band, while for the higher spectrum allocations the SBV solution should be implemented, as it overcomes full shared bandwidth usage. As shown in the following, this advantage rapidly increases with further expanded use of cable frequency spectrum beyond 35.2 MHz.

TABLE II. TONES AGGREGATION FOR COMPARISON PURPOSES BETWEEN NV AND SBV.

| Band number | 1 | 2 | 3 | 4 | 5 | 6 | 7 |
|---|---|---|---|---|---|---|---|
| Frequency range [MHz] | 0.138 – 3.75 | 5.2 – 8.5 | 14 – 17.66 | 17.66 – 22.08 | 22.08 – 26.50 | 26.50 – 30.90 | 30.90 – 35.20 |
| Bandwidth [MHz] | 3.612 | 3.30 | 3.66 | 4.42 | 4.42 | 4.42 | 4.30 |

*C. The fairness problem: how band partitioning can be sized*

In this Section we analyze the problem of band slot partitioning of such sub-channels to be adopted for SBV. Referring to Figure 1, the *i*-th generic sub-channel is partitioned into *K* consecutive band slots $B_i$, $i$ = 1,2,…,$K_i$ and $\sum_{k}^{P} B_i = B_{sc}$ where $B_{sc}$ is the sub-channel band. In principle, $B_i$ could be different and $K_i$ can vary with the sub-channel *i*. We only analyze the simplest case of constant $B_i = B$ for frequencies ranging from 35.2 MHz to 105.6 MHz. We consider uniform band slot partitioning by assigning to operators an equal number of consecutive tones in each band slot. Fairness should ensure that the two (or more) operators can provide (almost) the same data-rate to user(s) at a generic distance *d* from the street cabinet. In the considered case, fairness should be obtained by selecting the value of *B* so that the considered fairness criterion is met.

In the simple case of two operators only, fairness could be expressed in terms of the fraction of the data-rate difference $\Delta r_b$ as a function of the CAB-to-NT distance *d*, i.e.:

$$\Delta r_b(d) = \frac{|R_{b,OP1}(d) - R_{b,OP2}(d)|}{\max\{R_{b,OP1}(d), R_{b,OP2}(d)\}} \quad (6)$$

where $R_{b,OPi}(d)$, $i$ = 1,2 are the data-rates of the two operators at distance *d*. The band *B* should be selected so that $\Delta r_b(d) \leq \Delta_0$ for each $d \in (d_{min}, d_{max})$ and $\Delta_0$ is the target requirement and $d_{min}$, $d_{max}$ are the minimum and maximum distance where FttC service is expected to be deployed, respectively. The extension to three (or more) operators is not difficult and should be based on the maximum difference between the achievable data-rates among operators. In Figure 4 we plot $\Delta r_b$ as a function of the CAB-to-NT distance for different values of *B* and two upper frequencies, 35.2 MHz and 105.6 MHz. From Figure 4 it can be observed that, in general, the increase of band slot leads to the increase of $\Delta r_b$ above 5% due to the reduced granularity, i.e. fairness is not granted for each $d$. This leads to the conclusion that, to achieve an acceptable degree of fairness for each CAB-to-NT distance $d$, it is not possible to trivially partition the overall band above 17.6 MHz into $N_{op}$ consecutive band slots each one assigned to a single operator.[5]

---

[4] This assignment scheme guarantees the maximum achievable fairness degree among operators but it can be impractical when adjacent channel interference is considered.

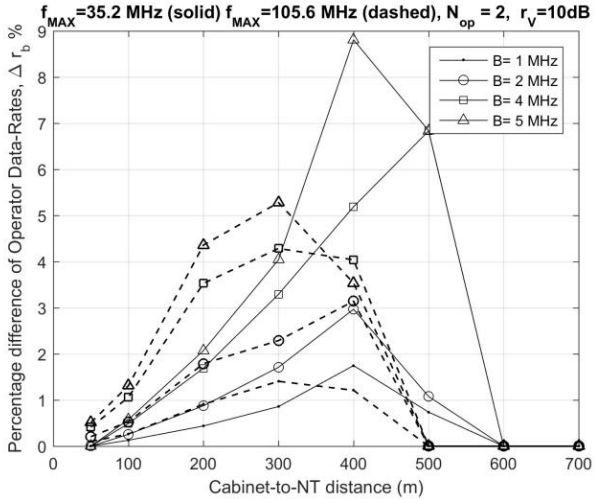

Fig. 4. Data-rate difference between operators vs distance for $f_{MAX}$= 35.2,105.6 MHz and variable band slot width.

To explain the reasons leading to the data-rate difference between operators, we observe that in our case in each band slot, lower tones have always been allocated to the first operator. With the increase of the $d$ the second operator using higher frequencies is slightly penalized in each band slot with respect to the other operator. This fact is better evidenced at large distances. These penalties cumulate and lead to the data-rate differences observed in Figure 4. The datarate difference between operators could be further reduced by swapping partitions between operators in each band slot i.e. in the first band slot the lowest partition is assigned to the first operator while in the second band slot the lowest partition is assigned to the second operator and so on, [12]. It is out of the scope of this paper to investigate an optimal DMT tone assignment in the band slot.

However, it should be observed that the existence of adjacent channel FEXT between consecutive partitions in the band slot(s) does not allow to select $B$ arbitrarily small.[6] At the same time the band slot dimension should be large enough to guarantee adjacent FEXT effects are confined to sub-carriers at the extremes of the adjacent partitions. To reduce adjacent[7]FEXT effects we can consider two solutions. The first consists in the introduction of guard bands between consecutive bandslot adjacent partitions. This lead to a reduction of the achievable data-rate. The second solution consists in adopting (some) power backoff mechanisms for sub-carriers at the extremes of the adjacent band-slot partitions and continue to use them for transmission even in the presence of adjacent FEXT interference. A hybrid approach consisting of both solutions could be also considered. In particular, guard bands (i.e. turn off of subcarriers at the borders of partitions) shall be considered only at higher frequencies where FEXT is important. For brevity, in this paper we do not further investigate on this aspect since the focus of this paper is on examining the potentials of the SBV and the considered simple band-slot partitioning strategy provides acceptable results.

Finally, as expected, data-rate differences rapidly drop to zero since the higher frequencies cannot be allocated with the increase of the CAB-to-NT distance and FEXT becomes negligible i.e. FEXT power falls below the background noise.

### D. Effects of the bandwidth increase

To evaluate the SBV performance as bandwidth increases, we have assumed the lower band (0–17.6 MHz) is nonvectored and we have imposed the fairness condition for frequencies above 17.6 MHz using the band partitioning technique detailed previously. The achievable aggregate data-rate per operator has been obtained by considering interference scenarios with high coupling FEXT conditions only considering both NV (see eq. (1)) and SBV (see eq. (2)), with NV in the same scenario described in Section III-B.

In Figure 5 we report the aggregate data-rate corresponding to the 10-th percentile as a function of CAB-to-NT distance $d$ for different values of the maximum frequency for $N_{op}$=2 (case $a$) and for $N_{op}$=3 (case $b$). Both the NV case and the SBV case have been considered. We assumed that the overall power to be used in the band above 17.6 MHz is 13.4 dBm. Following the considerations drawn in Section III, the aggregate datarate is the sum of the data-rate in the 0 − 17.6 MHz band under NV conditions with that obtained by applying SBV for frequencies higher than 17.6 MHz. The overall transmitted power, including that used in the 0 − 17.6 MHz is about 17 dBm. Computer calculations done with higher values of overall power showed that data-rate performance does not improve significantly beyond this value. Both medium load (i.e., $N_{us}$ = 12) and high load (i.e., $N_{us}$ = 24) scenarios have been considered. We extended the bandwidth up to a maximum frequency $f_{MAX} = 35.2 + n17.6$ MHz, $n = 0,...,4$. It is noted that for large distances (above 400 m) the datarate curves corresponding to different values of $f_{MAX}$ do not converge to a single curve. This is due to the flat transmission power spectrum assumption. In other words, having assumed that the transmission power beyond 17.6 MHz is constant, the increase in the number of tones with $f_{MAX}$ leads to a reduction of the maximum power available for each tone. Furthermore, for large distances higher frequencies become rapidly unusable and the saved power has not been re-allocated on the active tones. The variation of the SBV data-rate as the number of interfering users increases (see Figure 5) is due to the reduced performance in the 0–17.6 MHz band where NV strategy has been considered. A possible alternative could be including self-FEXT cancellation in the lower band which we did not consider in order to account for a worst case condition. As shown in Figure 5, the SBV technique provides significant performance improvement with respect to the NV case. Due to vectoring, SBV allows to exploit higher frequencies (even beyond 35.2 MHz) while this is not feasible under the NVonly condition. Figure 6 clearly outlines this important fact. In Figure 6 we show the achievable data-rate as a function of $f_{MAX}$ in the NV case and in the SBV case, considering both the medium load and the high load scenarios. Under the NV conditions the bands available at higher frequencies are unusable to improve data-rate even for relatively short CAB-to-NT distances. This is shown by the saturation effect due to FEXT that at high frequencies becomes dominant. Instead, thanks to vectoring SBV (including the NV data-rate contribution for frequencies below 17.6 MHz) can cope with FEXT and allows to significantly improve the overall data-rate achievable per operator as $f_{MAX}$ increases.

---

[5] This trivial frequency division rule requires to identify $N_{op}$−1 frequency values $f_i$ in the band above 17.6 MHz and to assign to the $l$-th operator the band between $f_{i-1}$ and $f_i$ where $f_0$ is the starting frequency of the higher band. In this case, achievable band slots are too large and fairness cannot be guaranteed.

[6] The case of $B$ equal to a single tone is a theoretical limit case where tones are assigned one at time and in an alternate manner to each operator. This is obviously impractical due to instabilities when using two or more separate SBV-DSLAM equipments.

[7] It should be remarked that adjacent FEXT interference is due to sidelobes of the sub-carrier spectrum. From preliminary experimental results in the 0–17.6 MHz it seems that adjacent FEXT interference is not of great concern at least for the considered frequencies.

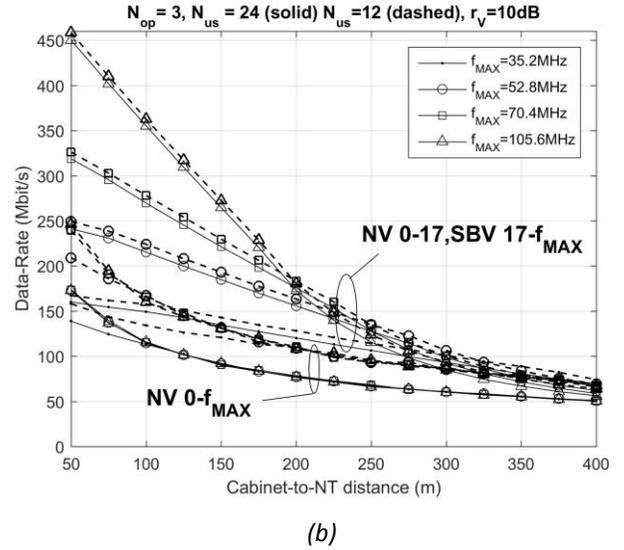

(b)

Fig. 5. 10-th percentile of the aggregate data-rate vs CAB-to-NT distance: $N_{op}$=2 (case *a*) and $N_{op}$=3 (case *b*), medium load and high load scenarios.

### E. Effects of vectoring imperfections

The effectiveness of the SBV technique is strongly related to the practical availability of vectoring algorithms able to ensure a significant level of FEXT reduction, aiming at achieving the AWGN condition. Ideally, the FEXT-free condition should be restored, so that the only impairment is the background noise. However, issues related to practical/economical implementation of the vectoring algorithm (fixed point algebra, channel estimation errors, non-ideal channel matrix inversion, etc.) may prevent ideal FEXT cancellation. To account for the unavoidable presence of residual FEXT after vectoring, we introduced in eq. (2) the degradation factor, $r_{V,k}$ > 1. It allows to express the residual FEXT as an increase of background noise at the *k*-th tone. In order to assess the effects of the residual

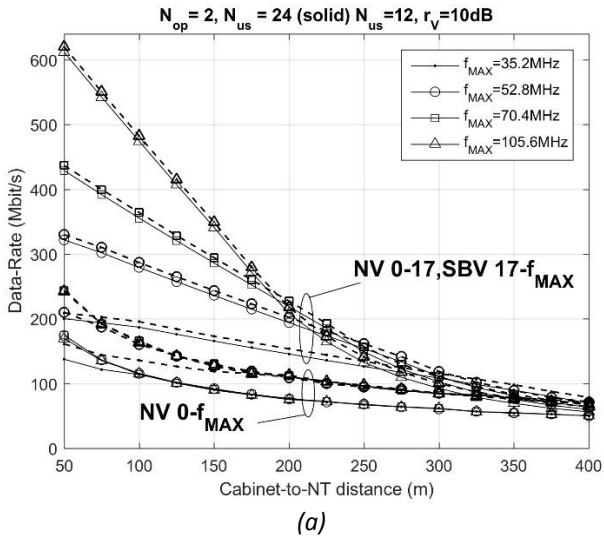

(a)

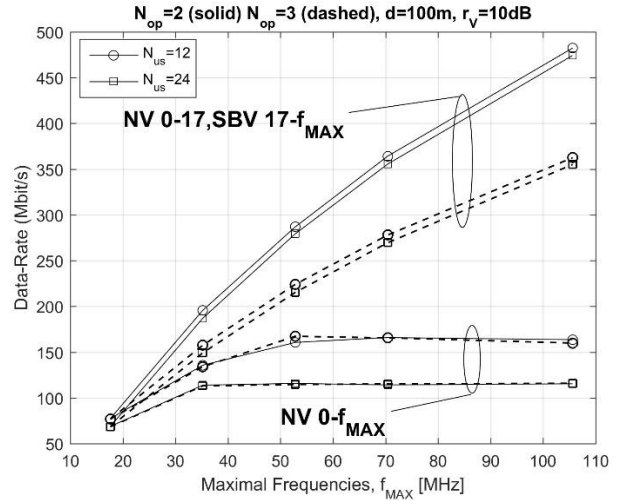

Fig. 6. Aggregate data-rate for the NV and SBV solutions as a function of $f_{MAX}$: $N_{op}$ = 2,3, CAB-to-NT distance d=100 m and medium and high load conditions.

FEXT degradation factor on the achievable SBV performance, in Figure 7 we show the achievable data-rate per band slot for $f_{MAX}$ = 35.2 MHz, and variable degradation factors ranging from 6 dB to 20 dB, with $N_{op}$ = 3 and under medium interference conditions. As shown, degradations in the order of 20 dB can have significant impact on the SBV performance, especially at relatively large CAB-to-NT distances (e.g., 250 m). For $r_V$ =20 dB constant for all tones, the NV outperforms the SBV of few Mbit/s for $f_{MAX}$ = 35.2 MHz. Nevertheless, for band slots at frequencies higher than 35.2 MHz even relatively large degradation can be tolerated because NV is unable to exploit higher frequencies and vectoring, even if degraded, is the only viable solution to exploit higher bands.

IV. COEXISTENCE WITH G.FAST

In addition to being fair between telecom operators, SBV is also a flexible approach. In fact, it allows independent usage of the bandwidth-distance resource by different operators, as well as complete freedom in technology selection. We need to account that operators naturally tend to move their fiber optic coverage closer to subscriber premises with time, possibly migrating to G.fast, which is able to provide increased datarate when vectoring is adopted. When SBV is used, the migration does not entail complex coordination procedures (see Figure 8). The key point here is that SBV makes easier coexistence between technologies and between operators in different scenarios. Even when one single operator only uses VDSL2 in FttC and FttDp configurations simultaneously, downlink power back-off is mandatory on the VDSL2/FTTDp to protect both systems [13]. Under similar conditions, when dealing with G.fast, the strong cross-talk couplings in such UBB systems make the cross-talk channel even stronger than the Dp-to-user direct link. According to [14], measurements show about 40 dB of flat SNR penalty from about 20 to

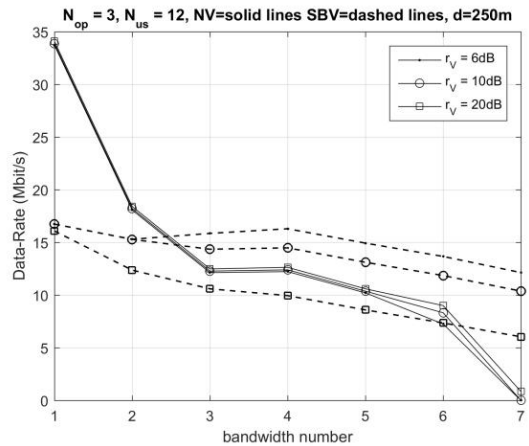

(b)

Fig. 7. Effects of Vectoring Degradation on SBV achievable performance for d=100 m in case (a) and for d=250 m in case (b).

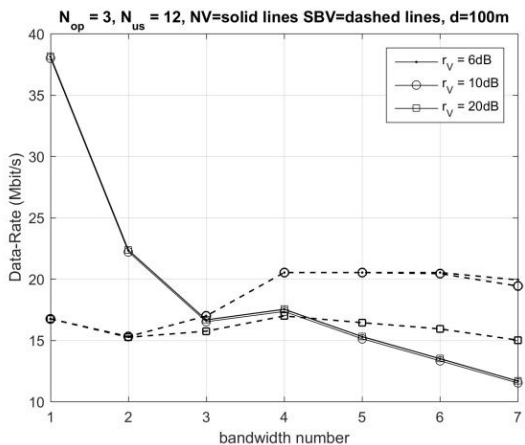

(a)

100 MHz when comparing conditions of absence of crosstalk cancellation with those with perfect vectoring. This is one main rationale for imposing vectoring in the G.fast (106a and 212a profiles) to eliminate self-FEXT.

Due to difficult coordination, coexistence between FttC and FttDp architectures can be a serious problem in the case of simultaneous use of mixed technologies in a multi-operator scenario. In a mixed technologies scenario, cross-talk between different services – i.e., VDSL2 (or e-VDSL) and G.fast – must be carefully considered. In fact, now alien cross-talk cannot be easily cancelled as is for self-FEXT, due to the different used technologies. Therefore, still dwelling on the simplest case of single operator conditions, even migration from VDSL2 to G.fast is not at all a trivial task.

One main issue is related to the time-division duplexing technique used for G.fast, instead of the frequency-division duplexing used for the DSL family. Being the G.fast 106a profile defined from 2.2 MHz to 106 MHz, while VDSL2 17a is from 25 kHz to 17.664 MHz, in general cross-talk power between the two systems can be FEXT or NEXT (near end cross-talk) and, differently for FEXT, vectoring is ineffective. For G.fast users, the most critical topology is when the VDSL2 street cabinet is close to the Dp where G.fast equipment is located, while the loop length for G.fast is long. For certain network topologies, when VDSL2 bands are occupied, G.fast data-rate improvement is marginally higher due to the heavy effect of interference mainly on the higher part of shared frequencies. The impact of G.fast on VDSL2 due to its smaller frequencies is even larger than the other way around, according to some simulation studies [15].

Obviously, when different telecom operators share the same cable, the problem is even more complex, as in general we cannot assume any kind of centralized control.

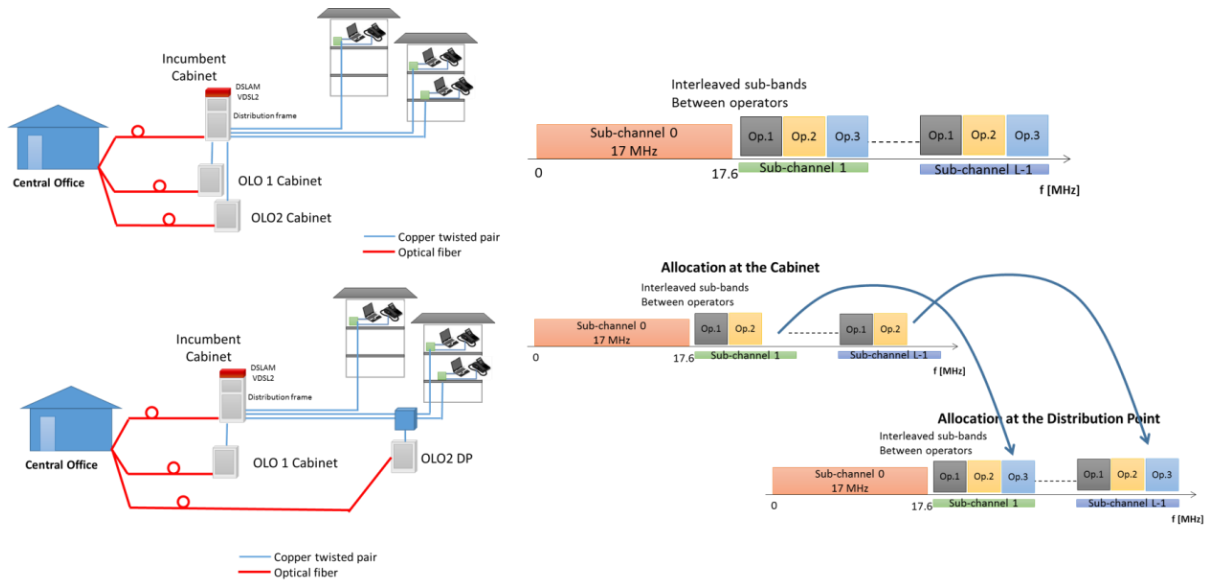

Fig. 8. Architecture evolution and band migration.

As future work, we intend to carry on interference analysis for the case of coexistence between G.fast (both 106a and 212a profiles) and e-VDSL (35.2 MHz) under multi-operator scenarios. However, the above-mentioned behaviors, already highlighted in the single operator case, allow affirming safely that bandwidth sharing should be avoided, not to jeopardize both e-VDSL and G.fast systems performance under realistic conditions. This conclusion is also in full agreement with those which led the standardization bodies to recommend that G.fast should only use the band above VDSL2. However, in case of e-VDSL an additional part of the precious lower frequencies band becomes unusable in a cable for G.fast unless the strategy depicted in Figure 8 is adopted.

## V. CONCLUSIONS

The need for data-rate increase due to bandwidth-hungry services for the residential customer is already forcing telecom operators to look for new solutions in the access architecture. Considering the regulatory constraints put forward to foster competition, vectoring can be discouraged in some European Countries. In this paper, we proposed a Sub-band Vectoring technique based on frequency division multiplexing to avoid alien-FEXT and on the use of vectoring to counteract selfFEXT, so that the SLU regulatory remedy is not sacrificed. In Countries where VDSL2 has been already deployed, the new eVDSL equipment can be introduced adopting vectoring in the extra bandwidth, only. It turns out that our SBV solution, in addition to providing fair resource sharing between operators, is quasi-optimal in terms of data-rate and does not present the well-known drawbacks of the MOV architectures studied so-far. Results of our calculations showed that SBV allows a data-rate per user up to 210 Mbit/s for e-VDSL with SBV and up to 620 Mbit/s for a bandwidth expanded up to 105.6 MHz, with two telecom operators. Finally, the frequency planning between operators can be extended to G.fast, during network evolution, so avoiding a similar interference problem between telecom operators that can be arbitrarily located at the street cabinet and/or at the buildings.